\documentclass[aps,amsmath,amssymb,12pt]{revtex4}

\usepackage{graphicx}
\usepackage{dcolumn}
\usepackage{bm}

\begin{document}
 
\title{Ultrafast nano-focusing with full optical waveform control}

\author{S. Berweger$^\dag$, J.M. Atkin$^\dag$, X.G. Xu, R.L. Olmon, and M.B. Raschke$^*$}
\affiliation{Department of Physics, Department of Chemistry, and JILA\\
University of Colorado at Boulder, Boulder, CO, 80309\\
*markus.raschke@colorado.edu}

\begin{abstract}
The spatial confinement and temporal control of an optical excitation on nanometer length scales and femtosecond time scales has been a long-standing challenge in optics.
It would provide spectroscopic access to the elementary optical excitations in matter on their natural length and time scales \cite{brixner05} and enable applications from ultrafast nano-opto-electronics to single molecule quantum coherent control \cite{assion98}.
Previous approaches have largely focused on using surface plasmon polariton (SPP) resonant nanostructures \cite{aeschlimann07} or SPP waveguides \cite{cao10,durach07} to generate nanometer localized excitations.
However, these implementations generally suffer from mode mismatch \cite{schuck05} between the far-field propagating light and the near-field confinement.
In addition, the spatial localization in itself may depend on the spectral phase and amplitude of the driving laser pulse thus limiting the degrees of freedom available to independently control the nano-optical waveform.
Here we utilize femtosecond broadband SPP coupling, by laterally chirped fan gratings, onto the shaft of a monolithic noble metal tip, leading to adiabatic SPP compression and localization at the tip apex \cite{stockman04,babadjanyan00}.
In combination with spectral pulse shaping \cite{weiner00, xu06} with feedback on the intrinsic nonlinear response of the tip apex, we demonstrate the continuous micro- to nano-scale self-similar mode matched transformation of the propagating femtosecond SPP field into a 20~nm spatially and 16~fs temporally confined light pulse at the tip apex.
Furthermore, with the essentially wavelength and phase independent 3D focusing mechanism we show the generation of arbitrary optical waveforms nanofocused at the tip. 
This unique femtosecond nano-torch with high nano-scale power delivery in free space and full spectral and temporal control opens the door for the extension of the powerful nonlinear and ultrafast vibrational and electronic
 spectroscopies to the nanoscale \cite{kubo05,bartels00}. 
\end{abstract}

\maketitle

In order to achieve the goal of an efficient nanometer confined femtosecond light source with \emph{independent} spatial localization and temporal control of the optical field, and which can freely be manipulated in 3D, the use of the unique properties of surface plasmon polaritons (SPP's) has long been discussed as a potential solution.
It is well established that the strong surface field localization and size and shape dependent resonances of SPP's as electromagnetic surface waves associated with collective charge density oscillations at noble metal-dielectric interfaces allow for sub-wavelength spatial control of even broadband optical fields \cite{schuller10}.
Elegant solutions to overcome the SPP diffraction limit \cite{yin05} and achieve
nano-focusing based on interference of localized SPP modes exist in the form of specially arranged cascaded, percolated, or self-similar chains of metal nanostructures as optical antennas \cite{volpe10,aeschlimann07,gunn10}.
However, the achievable optical waveforms at the nano-focus are often constrained by the phase relationship between the spectral modes already necessary to achieve the 3D nano-focusing \cite{durach07}.
This limits the degrees of freedom for full and structurally independent spatial and temporal control of the nano-focused field \cite{note}.
Related challenges persist for nano-devices in the form of tapered grooves, wires, or wedges \cite{volkov09,fang09,cao10}.
While they allow for nano-focusing via their propagating SPP waveguide properties
with favorable power transfer, scalability, and broad bandwidth,
many such geometries do not allow for full 3D spatial localization independent of spectral phase \cite{durach07, cao10}, and a substrate-based design \cite{volkov09} makes spatially and spectrally non-dispersive nano-focusing difficult.

In contrast, a 3D tapered tip as an SPP waveguide stands out due to its unique topology as a cone.
As has been proposed theoretically \cite{babadjanyan00,stockman04}, and recently demonstrated experimentally \cite{ropers07,neacsu10,sadiq11}, that geometry allows for true 3D focusing into a excitation volume as small as a few 10's of nm size.
The divergence of the effective index of refraction with decreasing cone radius experienced by an SPP propagating towards the apex leads to a continuous transformation of cylindrical modes and thus near adiabatic SPP nano-focusing into the apex of the tip.
\begin{figure}[t]
\includegraphics[width=.8\textwidth]{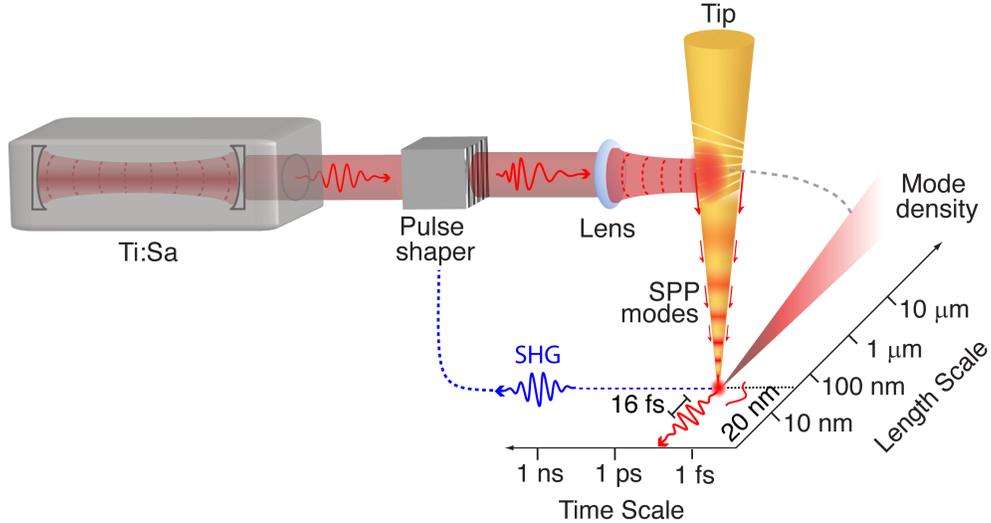}
\caption{
Micro- to nano-scale optical mode tranformation on a tip.
Broadband SPP coupling of a femtosecond laser pulse onto the shaft of conical Au tip is followed by adiabatic field compression into the nanoscale apex volume.
The spatio-temporal coordinate system depicts the associated mode confinement in space and time.
The SHG response from the tip is used as feedback for the pulse shaper to optimize pulse duration via MIIPS and to characterize arbitrary optical waveforms in the nanofocus via FROG or XFROG.}
\label{fig0}
\end{figure}

Despite some constraints on power transfer from the propagating SPP to the apex localized excitation due to SPP absorption and reflection,
this effect is only weakly wavelength-dependent and the nano-focusing mechanism is expected to be independent of spectral phase \cite{issa07}. 
Therefore, for sufficiently broadband SPP excitation, independent nanometer spatial and femtosecond temporal control should be achievable.  

As shown conceptually in Fig.~\ref{fig0} (for a detailed experimental description see supplementary material) for the experiment, laser pulses from a 10~fs Ti:Sa oscillator are passed through a pulse shaper for spectral amplitude and phase control \cite{weiner00}, 
and focused onto the shaft of a monolithic Au tip at 25$^\circ$ incidence using a long working distance objective.
The tips are formed through an electrochemical etching process, resulting in $\sim$10~nm apex radius, as discussed previously \cite{neacsu10}.
SPP's are launched onto the tip using a laterally chirped fan-shaped plasmonic grating element, allowing tunable broadband grating-coupling, in order to overcome the momentum mismatch between the incident wavevector and the SPP \cite{ropers07}.
The light emission from the tip apex is collected through a separate objective, with spatial and spectral filtering, and detected using a spectrometer with a N$_2$(l)-cooled CCD.
\begin{figure}[!ht]
\includegraphics[width=.8\textwidth]{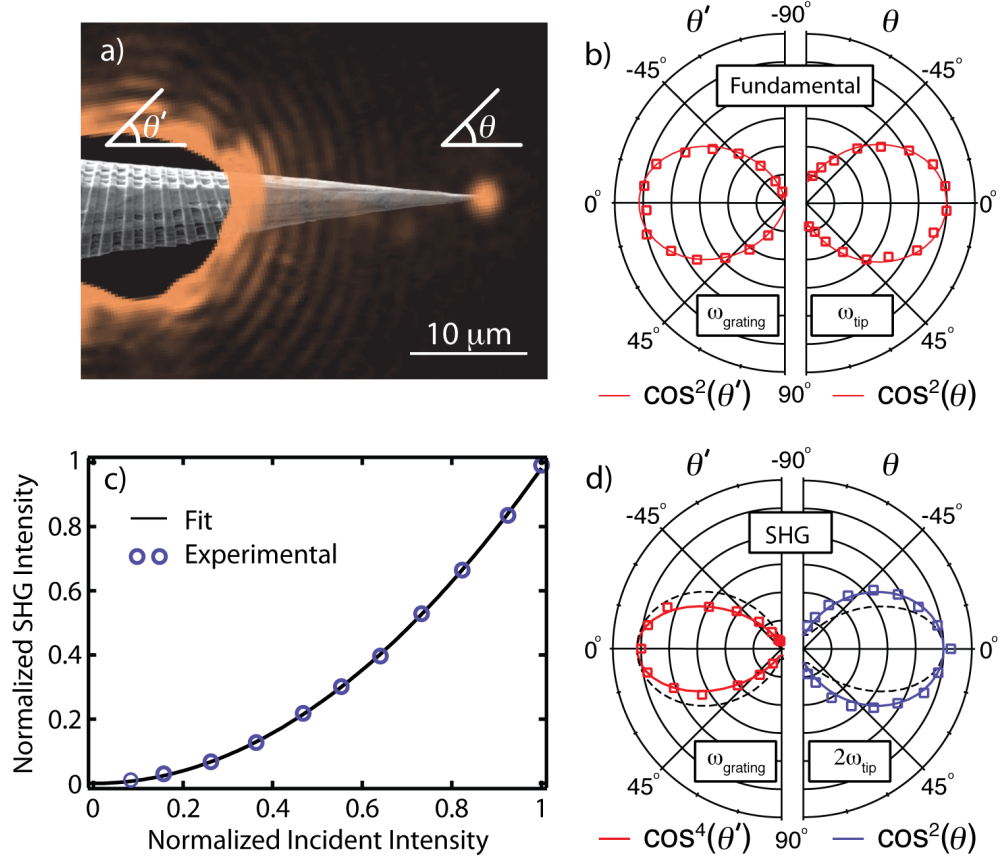}
\caption{Broadband tip nanofocusing and intrinsic nonlinear response: 
Optical image (a) of the grating coupled and apex emitted light superimposed with an SEM image of the tip with a broadband laterally chirped fan-shaped grating.
Polarization anisotropy (b) of the grating coupled and apex-emitted fundamental light, demonstrating the $\rm{cos}^2(\theta)$ dependence expected of a point dipole at the apex, oriented parallel to the tip axis \cite{neacsu10}.
Intensity dependence (c) of the apex emitted SHG on the incident light (blue circles) and a I$(2\omega)~\propto~\rm I (\omega)^2$ fit (black line).
Polarization anisotropy (d) of the apex emitted SHG, exhibiting a $\rm{cos}^2(\theta)$ dependent emission, and a $\rm{cos}^4(\theta)$ dependence on the grating incident fundamental light polarization.
}
\label{fig2}
\end{figure}

Fig.~\ref{fig2}a) shows an SEM image of a tip with a broadband grating superimposed with an optical image showing the far-field illumination incident on the grating and subsequent re-radiation of the spatially confined apex SPP field from a region $<$~25~nm in size, as demonstrated by scanning the tip over a nanometer edged reference sample \cite{neacsu10}.
The characteristic polarization anisotropy of both the apex-emitted fundamental and localized second-harmonic generation (SHG) light, arising from the broken axial symmetry at the apex \cite{neacsu05}, show the expected $\rm{cos}^2(\theta)$ intensity dependence expected for a point dipole emitter.
Note the different fundamental grating input polarization dependencies with $\rm{cos}^2(\theta)$ dependence for fundamental apex emission, and $\rm{cos}^4(\theta)$ dependence for SHG emission.
Typically, for $\sim$30~mW incident light on the grating we estimate the total apex emission to be $\sim$100~$\mu$W fundamental and $\sim$10~pW SHG.

After optimizing the grating illumination and coupling parameters for maximum coupling bandwidth or SHG intensity depending on application, the optical waveform of the nano-focus is controlled using spectral pulse shaping with SHG as the feedback parameter for the multiphoton intrapulse interference phase scan (MIIPS) algorithm used \cite{xu06}.
The MIIPS-optimized pulses are then characterized via interferometric frequency resolved optical gating (IFROG) \cite{anderson10} and the pulse transient is reconstructed with phase and amplitude information from the DC portion of the spectrogram using a standard FROG retrieval algorithm.
The pulse replicas as required for the IFROG autocorrelation measurements are themselves generated with the pulse shaper.
\begin{figure}[b]
\includegraphics[width=.5\textwidth]{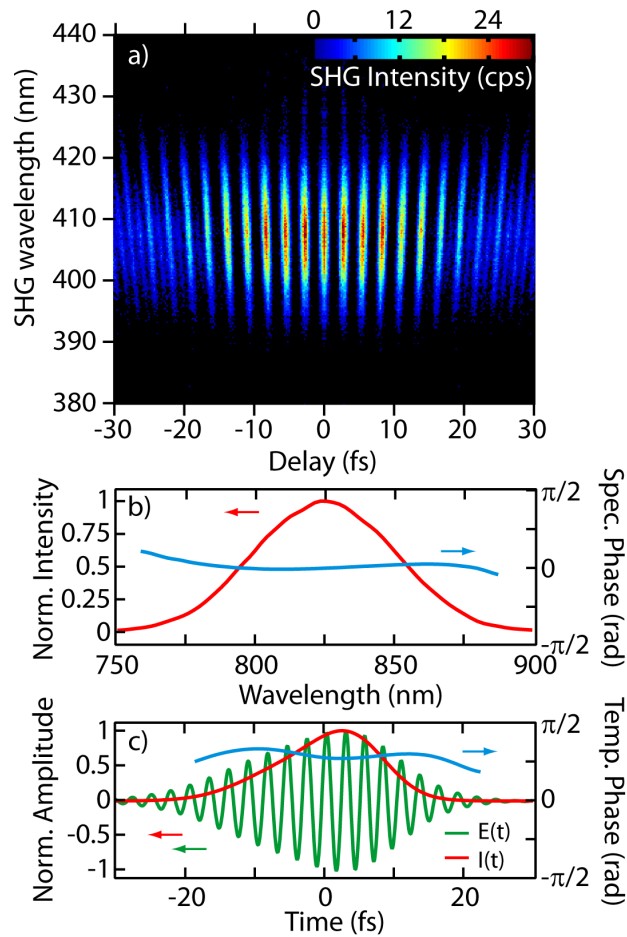}
\caption{
Nanofocusing of a few femtosecond pulse at the tip: 
FROG measurement based on apex localized SHG (a) of adiabatically nanofocused SPP.
MIIPS optimization for flat spectral phase yields a nearly transform limited 16 fs pulse for a 60 nm FWHM bandwidth (b and c) with reconstructed spectral and temporal phase (blue), intensity (red), as well as the reconstructed temporal electric field transient (green).
}
\label{fig3}
\end{figure}

Fig.~\ref{fig3}a) shows an IFROG trace for few-femtosecond broadband nanofocusing with corresponding reconstructed spectral (Fig.~\ref{fig3}b) and temporal (Fig.~\ref{fig3}c) intensity profiles (red) and phase (blue).
After MIIPS optimized flattening of the spectral phase to within 0.1~rad for the primary portion of the pulse, we obtain a transform-limited pulse with a duration of $\simeq$~16~fs  for the given coupling bandwidth of FWHM~$\simeq$~60~nm in this case.
The result is shown in Fig. 3c) along with the temporal electric field transient obtained (green).
\begin{figure}[b]
\includegraphics[width=.5\textwidth]{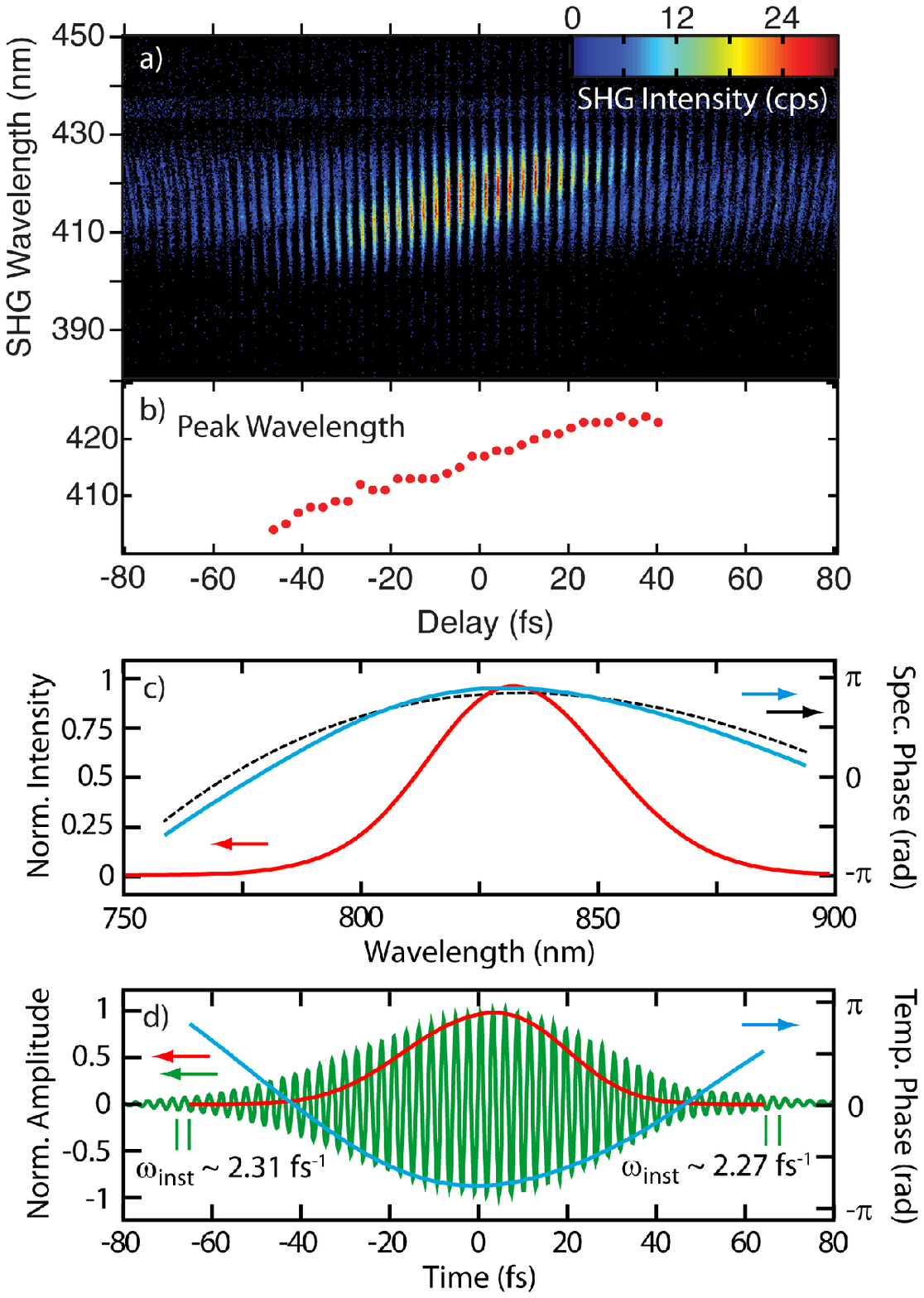}
\caption{
Deterministic arbitrary optical waveform control at the nanofocus:
XFROG measurement of the apex field with an applied group delay dispersion of 200~fs$^2$ with a transform limited 16 fs pulse as reference (a), with corresponding temporal variation in XFROG peak wavelength (b).
Reconstructed spectral characteristics (c) of the chirped pulse with intensity (red), and spectral phase (blue) in comparison with the theoretical 200 fs$^2$ GDD applied (dashed).
Reconstructed chirped electric field transient (green) (d).
}
\label{fig4}
\end{figure}

To demonstrate the capability to nanofocus a femtosecond pulse of arbitrary waveform we generate a chirped pulse with a group delay dispersion of $\phi_2=200$ fs$^2$ at the tip apex as an example. 
Fig.~\ref{fig4}a) shows the measured XFROG trace using a transform limited 16 fs pulse with identical spectrum as reference.
Fig.~\ref{fig4}b) shows the associated temporal variation in instantaneous carrier wavelength.
The reconstructed spectral intensity of the chirped pulse is shown in Fig.~\ref{fig4}c) (red).
The comparison and good agreement between the reconstructed phase (solid blue) and the applied phase function of $\phi_2=200$ fs$^2$ (black dashed) as set by the pulse shaper show the high degree of accuracy of this procedure.
Shown in Fig.~\ref{fig4}d) are the corresponding time-domain intensity (red), phase (blue), and electric field transient (green) with the shift in the instantaneous frequency indicated across the chirped pulse.

These two examples demonstrate the unique ability of the 3D tapered tip for simultaneous SPP nanofocusing and optical waveform control compared to the alternative geometries and approaches mentioned above.
By relying on the diverging index of refraction experienced by propagating SPP's with decreasing cone radius, the mode transformation into a nanoscale excitation at the tip apex remains continuous and impedance matched \cite{stockman04}, while the uniform taper surface prevents the otherwise typical scattering and reflection losses at structural discontinuities.
Due to the radial symmetry and decreasing SPP group velocity, full nanoconfinement in all spatial dimensions is achieved \cite{neacsu10,berweger10,sadiq11}.

In addition, the tips allow for the desired broadband excitation necessary for true femtosecond optical control. 
As predicted theoretically, the nanofocusing efficiency is maximal in the 800~nm range for a tip cone angle of $\sim$14$^\circ$ \cite{issa07} as used in our experiment with previously demonstrated monochromatic nanofocusing efficiencies of up to 9$\%$ \cite{berweger10}.
Since the SPP nanofocusing mechanism does not rely on localized SPP resonance conditions, its weak wavelength dependence provides phase-independent nanofocusing over a broad wavelength range as desired for ultrafast pulses as short as just a few femtoseconds \cite{issa07}.
As a direct consequence, the spectral phase and amplitude of the coupled pulse are retained as degrees of freedom that can be controlled independently.

In contrast to other nano-focusing structures based either on localized plasmon resonances \cite{aeschlimann07} or propagating SPP modes \cite{durach07,cao10} which exhibit a dependence of the
spatial localization on the spectral and phase characteristics of the excitation field, we control the pulse duration and optical waveform at the tip apex via {\em deterministic} pulse shaping, as opposed to {\em adaptive} techniques, thus reducing the computation duration of the optimization procedure as well as algorithm complexity.
Furthermore, in order to characterize the designed nanofocus waveform both in terms of spectral phase and amplitude, 
the nanofocused apex field conveniently generates local SHG due to the broken axial symmetry of the apex \cite{neacsu05}.
This structurally intrinsic coherent nonlinear response thus enables a direct means to apply appropriate spectrogram-based nonlinear wavemixing pulse characterization techniques (e.g., FROG, XFROG) for the complete optical waveform determination and optimization without the requirement for a separate nonlinear material or asymmetric structure for nonlinear optical frequency conversion.

The theoretical limit for the shortest attainable pulse duration at the tip apex is determined only by the coupling bandwidth and the dephasing time $T_2$ in the case of spectral overlap with a localized SPP of the tip apex.
While the tips used in our off-resonant experiments shown here typically exhibit SPP resonances in the vicinity of 650~nm, the use of localized tip SPP's with measured dephasing times of up to $T_2 \simeq 20$ fs \cite{anderson10} can provide additional capability for tailoring the optical waveform with higher field enhancement, albeit at the expense of minimal pulse duration.

The details of the nanofocusing mechanism and the degree of adiabaticity are not yet completely understood as the optimal nanofocusing conditions require minimization of both SPP propagation damping and reflection losses, which increase and decrease with adiabatic conditions, respectively.
This includes the propagation induced dispersion which is of particular interest for femtosecond nanofocusing. 
However, considering a group velocity dispersion of $\partial^2k/\partial \omega^2 \simeq$ 0.4~fs$^2/\mu$m for an SPP propagating on a flat surface, the dispersion introduced by a propagation length of 20~$\mu$m is expected to be small. 
This is in good agreement with comparison of grating-coupled and free-space MIIPS measurements showing little dispersion introduced by grating-coupling and subsequent SPP propagation and nanofocusing, although some tip-to-tip variability is observed. 
While arbitrary dispersion could be compensated within the capability of the pulse shaper using MIIPS, the apparent low dispersion provides favorable conditions for pulse optimization and characterization by yielding already large initial SHG levels.

The use of a grating as coupling element can provide high coupling efficiencies, but some scattering losses are inevitable.
The fan-shaped gratings developed for this work enable a high coupling bandwidth with FWHM of up to $\sim 100$ nm (see supplementary material).
Improvements in the form of grating structures with optimized groove depth and holographic geometries, chirped gratings including Bragg reflectors \cite{lopez07}, broadband coupling via fabricated micro-prism onto the tip shaft \cite{sanchez02}, spatial pulse shaping in the far-field excitation focus \cite{piestun01}, and tip fabrication with reduced surface roughness could further improve the tip performance for nanofocus optical waveform control. 
Without the need for resonance behavior in order for spatial localization to occur, the process can be extended to a wide range of wavelengths, limited in principle only by material damping at short wavelengths, and taper angle and reduced spatial field confinement at long wavelengths.

In summary, we have demonstrated independent nanometer spatial and femtosecond temporal optical waveform control, enabled by nanoscale field concentration via adiabatic SPP nanofocusing into monolithic gold tips, which is intrinsically broadband and independent of the instantaneous frequency and spectral phase of the excitation field.
The in principle impedance matched far-field to near-field mode transformation allows for efficient power transfer into a nanoconfined volume at the tip apex. 
This light source with arbitrary waveform control at the nanoscale is of a fundamentally new quality compared to both conventional far- and near-field sources.
It allows for the systematic extension of near background-free scanning probe microscopy \cite{berweger10,deangelis09,sadiq11} to the nanoscale implementation of many forms of nonlinear and ultrafast spectroscopies for spatio-temporal imaging \cite{terada10}.
This offers all-optical access to the study of nonequilibrium carrier and lattice excitations and their correlations on the level of their natural femtosecond time and nanometer length scales, thus providing unprecedented microscopic insight into the origin of complex biological, organic, or correlated electron materials.
It allows for quantum coherent control of chemical reactions \cite{assion98} on the nanoscale, quantum information processing, provides a tool for nano-photonic circuit analysis, and, with the high field compression, new avenues for extreme nonlinear optics such as higher harmonic generation \cite{bartels00} or femtosecond electron pulse generation \cite{ropers07a}.

\section*{Acknowledgements}
We would like to acknowledge valuable discussions with Mark Stockman, Christoph Lienau, and Hrvoje Petek as well as funding from the National Science Foundation (NSF CAREER grant CHE 0748226).

\section*{Author contributions}
S.B., X.G.X., and M.B.R. conceived the experiments as carried out and analyzed by S.B. and J.M.A. using the pulse shaper designed and built by X.G.X.;
R.L.O. and S.B. designed the grating tips which were fabricated by R.L.O.;
S.B., J.M.A., and M.B. R. prepared the manuscript.
$\dag$ S.B. and J.M.A. contributed equally to this work.

\bibliography{grating_SHG.bib}

\begin{thebibliography}{32}
\bibitem[{not()}]{note}
\bibinfo{note}{The solution in the form of the emergent nano-plasmonic laser
  (surface plasmon amplification by stimulated emission of radiation, SPASER)
  can provide the desired highly localized power, but does not yet provide the
  required broad band gain and phase control for pulsed operation}.
\end{thebibliography}

\end{document}


\title{Supplementary Material for:\\
Ultrafast nano-focusing with full optical waveform control}

\author{S. Berweger$^\dag$, J.M. Atkin$^\dag$, X.G. Xu, R.L. Olmon, and M.B. Raschke$^*$}
\affiliation{Department of Physics, and Department of Chemistry, and JILA \\
University of Colorado at Boulder, Boulder, CO, 80309\\
*markus.raschke@colorado.edu}

\maketitle

\section*{Experimental}

\begin{figure}[t]
\includegraphics[width=.8\textwidth]{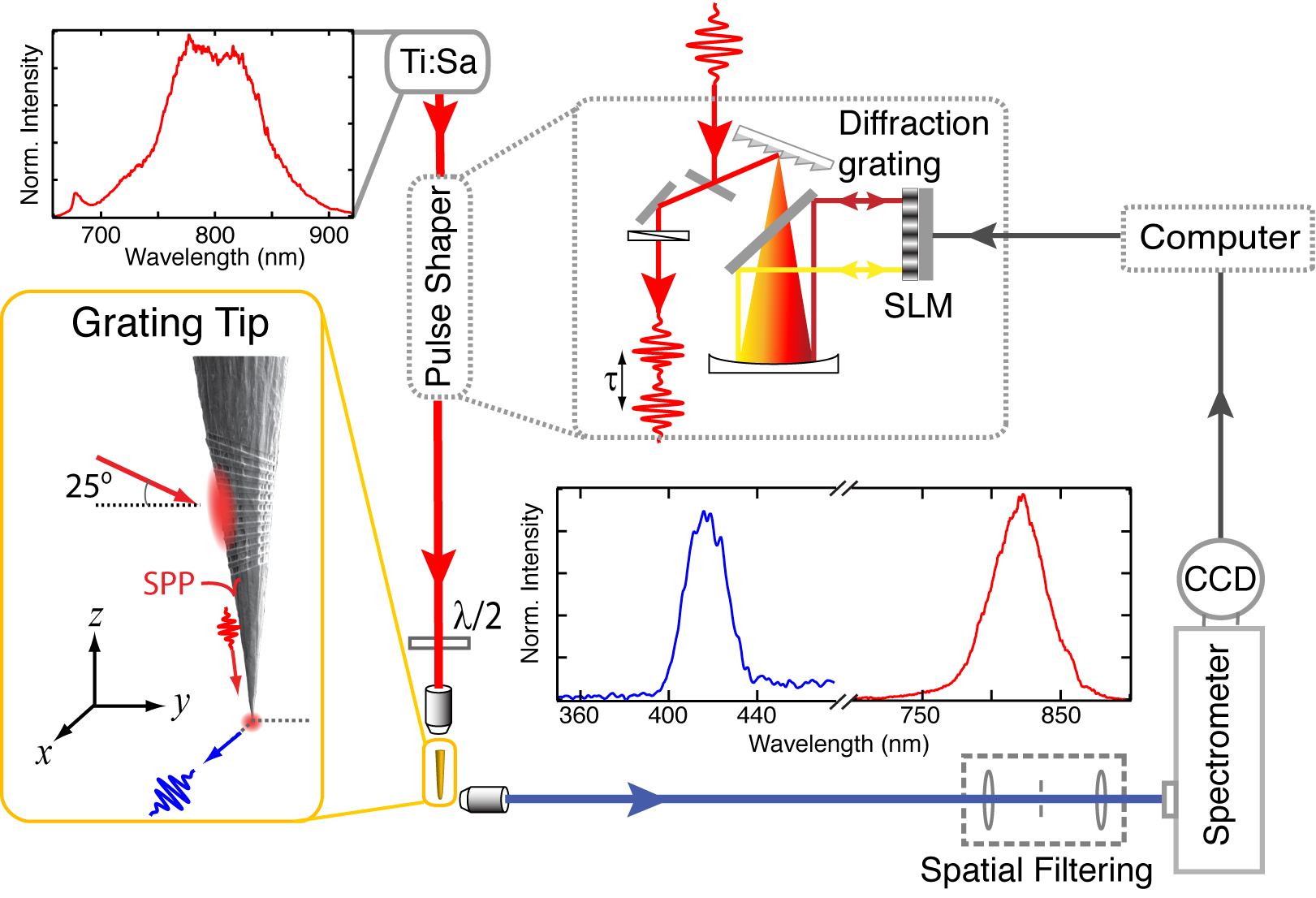}
\caption{
Schematic of the experimental setup consisting of Ti:Sa laser, pulse shaper, and tip illumination and nano-focused apex emission detection optics. 
}
\label{supp1}
\end{figure}

The schematic of the experimental setup is shown in Fig.~\ref{supp1}.
For our light source we use pulses from a Ti:Sa oscillator with center wavelength $\sim 800$ nm, bandwidth FWHM $\sim 100$ nm, nominal pulse duration $\sim$10~fs, pulse energy $\sim$8~nJ, and repetition rate 75~MHz (Femtolasers Synergy).
The pulses are passed through a pulse shaper using a folded 4f geometry with a 640-pixel liquid crystal spatial light modulator (CRI Inc.) for independent spectral amplitude and phase control.
Pulses are focused onto the shaft of electrochemically etched Au tips \cite{ren04} at 25$^\circ$ incidence using a long working distance objective with NA~=~0.35, focal length = 20.5 mm, focus size $\simeq$~8~$\mu$m$^2$, and fluence $<$~6$\times$10$^{5}$~W/cm$^2$ (Nikon SLWD).
A plasmonic grating, cut by focused ion beam milling, enables the launching of propagating surface plasmon polaritons (SPP's) onto the tip where they are concentrated under near-adiabatic conditions into a region 10's of nm in size at the tip apex where a portion of the fundamental light is re-emitted \cite{ropers07,neacsu10}. In order to provide continuously adjustable nanofocused power a $\lambda$/2 plate is used to change the grating coupling efficiency. 

In addition to the emitted fundamental light, the broken axial symmetry at the tip apex allows for the local generation of optical second harmonic (SHG) \cite{neacsu05}.
The apex emitted fundamental and SHG light is collected through a separate objective with NA~=~0.5 and k-vector direction 90$^\circ$ azimuthal with respect to illumination.
The short SPP propagation lengths on Au of $1/\Im(2k_{\rm{SPP}})$~=~300~nm at $\lambda$~=~400~nm ensures detection of apex-localized emission.
The light is then spectrally filtered and detected using a spectrometer with a N$_2$(l)-cooled CCD (Princeton Instruments).

Since the nanofocusing process is independent of the spectral phase of the pulses, this allows for the deterministic optimization and control of the nanofocused pulse characteristics.
Thus, by measurement of the spectral phase through the $\chi^{(2)}$ SHG response, we can generate transform-limited pulses by flattening the spectral phase using a multiphoton intrapulse interference phase scan (MIIPS) algorithm \cite{xu06}.
Furthermore, full pulse characterization can be achieved though interferometric frequency resolved optical gating (IFROG) \cite{trebino97,stibenz05,anderson10} from reconstruction of the DC portion of the interferogram using a standard FROG retrieval algorithm.
In order to obtain reliable IFROG measurements identical coupling conditions for both pulse replicas must be ensured.
This can be accomplished using the pulse shaper to generate perfectly collinear pulse pairs with the applied mask $M(\omega)$:
\begin{equation}
M(\omega)=\rm{cos}(\phi(\omega)/2)e^{i\phi(\omega)/2}
\label{eq1}
\end{equation}
with carrier frequency $\omega$ and spectral phase $\phi$, where
\begin{equation}
\phi(\omega)=\phi_0+\phi_1 (\omega-\omega_0)+\phi_2 (\omega-\omega_0)^2/2!+...
\label{eq2}
\end{equation}
resulting in two pulse replica one of which is delayed in time by $\phi_1$ \cite{shim06}.
Furthermore, with the capability to generate arbitrary (limited only by resolution of the spatial light modulator) relative $n$-th order spectral phase (dispersion) $\phi_n$ between the pulses, XFROG measurements can also be performed.

\section*{Fan Grating}

\begin{figure}[b]
\includegraphics[width=.6\textwidth]{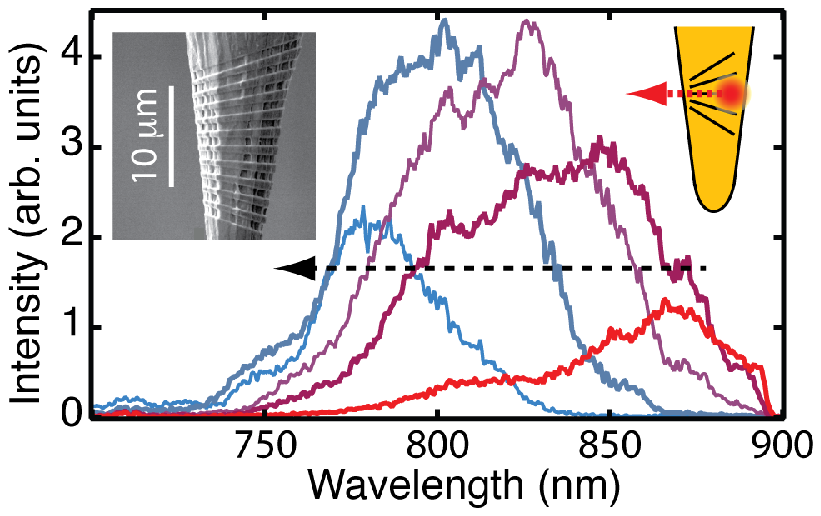}
\caption{
Grating coupling spectra acquired with the illuminating focus scanned across the fan-shaped grating in a horizontal direction to demonstrate the variation in the coupling wavelength.
Inset: SEM image of a fan-shaped grating fabricated on the shaft of a Au tip via FIB.
}
\label{supp2}
\end{figure}
The plasmonic grating is utilized to overcome the momentum mismatch between the free-space propagating light with in-plane momentum $k_{in}$ and the SPP with momentum $k_{SPP}$~=~$k_{in}$~+~$\nu G$ with the reciprocal lattice vector $G$~=~$2\pi/a_o$, integer $\nu$, and the grating period $a_o$ \cite{raether,neacsu10}.
However,
even for regular gratings, the grating coupling conditions typically depend sensitively on structural features smaller in size than the resolution of $\sim$30~nm of the focused ion beam (FIB) mill used to cut the grating \cite{renger07,ditlbacher03} and some variation in grating coupling bandwidth is seen between different tips.
Since the peak coupling wavelength depends only on $a_o$ and the angle of the incident light via $k_{in}$, 
adjustments in the incident angle through mechanical tilting of the tip must be used to optimize coupling conditions.

In order to overcome limitations imposed by conventional plasmonic grating design we utilize specially designed gratings varying $a_o$ horizontally, resulting in a fan shape, to enable both enhanced broadband coupling and spectral tuning.
An SEM image of the fan-shaped grating utilized for this purpose is shown in the inset of Fig.~\ref{supp2}.
With a varying grating period across the incident focus, a broader coupling spectrum can be achieved for suitable illumination conditions.
While the coupling conditions 
vary between different tips, the coupling bandwidth is generally improved utilizing fan-gratings and coupling spectra with FWHM up to $\sim$100~nm can be achieved.
Additionally, this provides the benefit of allowing for easy tuning of the coupling wavelength merely by adjusting the lateral position with a tight illumination focus.
As shown in Fig.~\ref{supp2}, upon shifting the illuminating focus across the grating from a region of longer grating period to one of shorter period the coupling spectrum is blue-shifted as expected.

Alternative approaches to broadband gratings include the use of gratings chirped along the tip axis.
However, in order to avoid reflection and thus replication of pulses \cite{lopez07} from the distal portions of the grating, such structures have not been used for the present studies.
Also, single groves could be used, providing broad-band coupling albeit at the expense of coupling efficiency.

\bibliography{grating_SHG.bib}